\documentclass[journal]{IEEEtran}

\usepackage{amsmath}
\usepackage{algorithm}
\usepackage{algorithmic}
\usepackage{booktabs}
\usepackage{multirow}
\usepackage{cite}
\usepackage[utf8]{inputenc} 
\usepackage[T1]{fontenc}    
\usepackage{hyperref}       
\usepackage{url}            
\usepackage{booktabs}       
\usepackage{amsfonts}       
\usepackage{nicefrac}       
\usepackage{microtype}      
\usepackage{xcolor}         
\usepackage{graphicx}
\usepackage{float}
\usepackage[numbers,sort&compress]{natbib}
\usepackage{caption}

\newfloat{figtab}{htb}{fgtb}
\makeatletter
  \newcommand\figcaption{\def\@captype{figure}\caption}
  \newcommand\tabcaption{\def\@captype{table}\caption}
\makeatother

\def\BibTeX{{\rm B\kern-.05em{\sc i\kern-.025em b}\kern-.08em
    T\kern-.1667em\lower.7ex\hbox{E}\kern-.125emX}}

\title{SASLO: A Scene-Aware Spatial Layout Optimization System for AR-SSVEP
}
\author{
\IEEEauthorblockN{Beining Cao\textsuperscript{1}, Xiaowei Jiang\textsuperscript{1}, Charlie Li-Ting Tsai\textsuperscript{1}, Daniel Leong\textsuperscript{1}, Thomas Do\textsuperscript{1}, Chin-Teng Lin\textsuperscript{1}\textsuperscript{*}}\\

\IEEEauthorblockA{
\textsuperscript{1} Australian AI Institute, School of Computer Science,\\ Faculty of Engineering and Information Technology, University of Technology Sydney}

\thanks{\textsuperscript{*}Corresponding author: Chin-Teng Lin. Email: Chin-Teng.Lin@uts.edu.au}

}

\begin{document}
\maketitle

\begin{abstract}
Steady-state visual evoked potential (SSVEP) is widely used in brain-computer interfaces (BCIs) due to its high reliability. By integrating augmented reality (AR), AR-SSVEP embeds visual stimuli into real-world environments, enabling more intuitive interaction. However, AR-SSVEP is subject to real-world scene factors, such as high background luminance, which reduce stimulus saliency and weaken SSVEP elicitation. Adaptive adjustment of the stimulus layout is a promising approach to mitigating adverse scene-induced effects. However, previous studies have primarily focused on offline analyses of scene-related factors under indoor conditions, whereas research on real-time, scene-adaptive optimization for outdoor AR-SSVEP remains limited.

Therefore, a scene-aware spatial layout optimization (SASLO) system for AR-SSVEP is proposed, which jointly considers scene luminance and inter-stimulus distance (ISD) for adaptive stimulus layout optimization. Scene luminance is estimated with an RGB-CIE-based method, and the extracted context is fed into a linear contextual bandit (LCB) model to recommend optimized spatial layouts. Two pilot single-factor experiments were conducted to investigate the effects of luminance and ISD on SSVEP performance and to construct reliable rewards for model training. The proposed joint optimization method was further validated in an outdoor online experiment involving ten subjects, achieving an average accuracy of 0.89 and an information transfer rate of 35.74~bits/min with a 3~s input window, and consistently outperforming two baseline methods. Overall, the proposed SASLO system is shown to improve the performance of AR-SSVEP in real-world outdoor environments.

\end{abstract}


\section{Introduction}

\IEEEPARstart{E}{lectroencephalography} (EEG)-based steady-state visual evoked potential (SSVEP) is a well-established brain-computer interfaces (BCIs) paradigm with high SNR and reliable frequency-tagging characteristics, requiring minimal calibration \cite{wei2025attention, zhao2016ssvep}. In SSVEP paradigms, periodic neural responses synchronized with the stimulus frequency and its harmonics are elicited in the visual cortex when subjects fixate on visual stimuli modulated at specific frequencies \cite{wang2022stimulus}. As a result, hands-free interaction is enabled by decoding user intent from frequency-specific SSVEP responses elicited by visual fixation on different stimuli \cite{cao2025emd,  lin2020direct}. Conventional computer screen-based SSVEP (CS-SSVEP) systems achieve high decoding accuracy in controlled laboratory settings \cite{hamidi2024optimization, liu2022ssvep}. Users are required to shift their attention between the stimulus display and the intended target, disrupting natural perception-action coupling and limiting real-world applicability.
To overcome these constraints, augmented reality-based SSVEP (AR-SSVEP) systems have been introduced, in which SSVEP stimuli are integrated into real-world scenes through optical see-through (OST) head-mounted displays \cite{du2022visual, ke2025dataset}. This unified visual presentation enables the concurrent perception of virtual stimuli and physical objects, supporting more natural interaction in AR environments \cite{park2022brain}. 

Nevertheless, AR-SSVEP exhibits lower accuracy than CS-SSVEP, mainly due to interference from real-world environments \cite{si2018towards,ravi2022enhanced, zhang2022ambient, do2020estimating, zhang2023improving, kim2025performance, do2021retrosplenial, angrisani2023wearable}. In AR settings, background scene characteristics substantially affect stimulus perception and the elicited SSVEP responses. Previous studies have mainly investigated the effects of background-related factors, such as color \cite{kim2025performance}, luminance \cite{zhang2023improving} and background motion \cite{shu2015visual} on AR-SSVEP. Among these studies, color is the only factor that has been investigated for online optimization. Specifically, Kim et al. investigated adaptive color-contrast modulation in AR-SSVEP systems and reported that color-based contrast enhancement led to a modest improvement in decoding accuracy \cite{kim2025performance}.

In addition to color modulation, the spatial layout of stimuli affects the background luminance at the stimulus locations, which further influences stimulus saliency and the elicited SSVEP responses. In real-world environments, background luminance varies across locations, causing stimuli at different positions to elicit SSVEP responses of varying strength. Stimulus saliency can be strengthened by placing stimuli in regions with lower luminance \cite{zhang2022ambient, reitelbach2024optimal}. Consequently, optimizing the spatial layout allows stimulus positions to be adjusted to control local background luminance, leading to additional performance benefits. 
On the other hand, in typical SSVEP paradigms, multiple stimuli are presented simultaneously, and spatial layout optimization based solely on background luminance can easily result in densely arranged layouts. It has been demonstrated that inter-stimulus distance (ISD) plays a critical role in SSVEP elicitation \cite{ravi2019user,zhao2020ssvep}. Visual competition can be induced by dense stimulus layouts \cite{ng2011effect}, resulting in mutual interference and weakened SSVEP responses. Even when individual stimuli are locally optimized based on background luminance, the overall configuration may remain suboptimal if the spatial spacing among stimuli is not adequately considered. Therefore, in addition to luminance-based optimization, ISD is also a key factor in spatial layout design. While prior studies have conducted preliminary explorations of these factors \cite{zhao2020ssvep, zhang2022ambient, ravi2019user}, the analyses have mainly been performed offline or under controlled indoor conditions. As a result, online scene-aware spatial layout optimization is required for robust AR-SSVEP operation in outdoor environments.

For online spatial layout optimization, a scene-aware strategy is required to automatically adapt stimulus layouts for improved AR-SSVEP robustness. In this task, scene perception of luminance and spatial parameter recommendation are essential. For scene perception, visual luminance can be estimated from RGB images through color space transformations \cite{fairchild2013color}. Specifically, transformations to the Commission Internationale de l’Éclairage (CIE) color space are widely used to approximate perceptual luminance by accounting for the unequal spectral sensitivity of the human visual system \cite{briggs2020colour}. Color space transformation-based luminance estimation methods are computationally efficient and suitable for online systems. By contrast, deep learning-based models achieve high accuracy and spatially consistent luminance estimation under complex lighting conditions \cite{cai2023retinexformer, bolduc2023beyond}.
However, the applicability of these methods to real-time on-device AR processing is limited by their higher computational cost.

With the estimated luminance context as input, a recommendation model is necessary to adaptively optimize the stimulus layout. Within the research on recommendation systems, contextual bandits (CB) algorithms are commonly adopted as a lightweight framework for context-dependent decision making with an exploration-exploitation trade-off~\cite{li2010contextual}. In a CB framework, contextual features are taken as inputs, candidate decisions are represented as ‘arms’ and selected as outputs, and a scalar reward is observed to guide policy updates. CB methods are commonly categorized into linear contextual bandits (LCB) \cite{agrawal2016linear} and neural contextual bandits (NCB) \cite{zhou2020neural}.
LCB typically model the context–reward relationship using linear assumptions, enabling efficient learning through linear function approximation.
In contrast, NCB employ neural networks to capture more complex, nonlinear context–reward relationships. Decisions are made by estimating expected rewards conditioned on contextual input. In this work, CB-based methods are well-suited, as extracted scene information naturally serves as context and multiple perceptual constraints can be jointly encoded in the reward function.

Therefore, this work proposes a scene-aware spatial layout optimization (SASLO) system for AR-SSVEP applications. Specifically, an SSVEP-based object selection task \cite{chen2020ssvep} is adopted as the experimental application, in which subjects select the target object by attending to the associated SSVEP stimulus. Scene perception is realized through an RGB-based luminance estimation method, and a LCB model is employed for layout recommendation. In the proposed system, scene luminance and ISD are jointly considered for optimization, enabling adaptive recommendation of the optimal spatial layout of stimuli under varying scenes. To obtain realistic reward signals for model training, offline experiments were conducted to separately evaluate the effects of luminance and ISD on SSVEP responses, from which human-derived rewards were obtained. Based on the LCB model, an online AR-SSVEP system was further implemented and evaluated in outdoor environments. Comparative evaluations demonstrate that the proposed joint optimization-based SASLO system outperforms both the luminance-only optimization and the non-optimization baseline in an AR-SSVEP-based object selection task.

The contributions of this study are summarized as follows:
\begin{enumerate}
\item A novel scene-aware spatial layout optimization method for AR-SSVEP is designed, jointly considering luminance and ISD.
\item The recommendation model is trained using human-derived rewards obtained from offline experiments.
\item An online SASLO system is developed and its effectiveness is validated under outdoor conditions.

\end{enumerate}

\begin{figure*}[h!]
    \centering
    \includegraphics[width=1\linewidth]{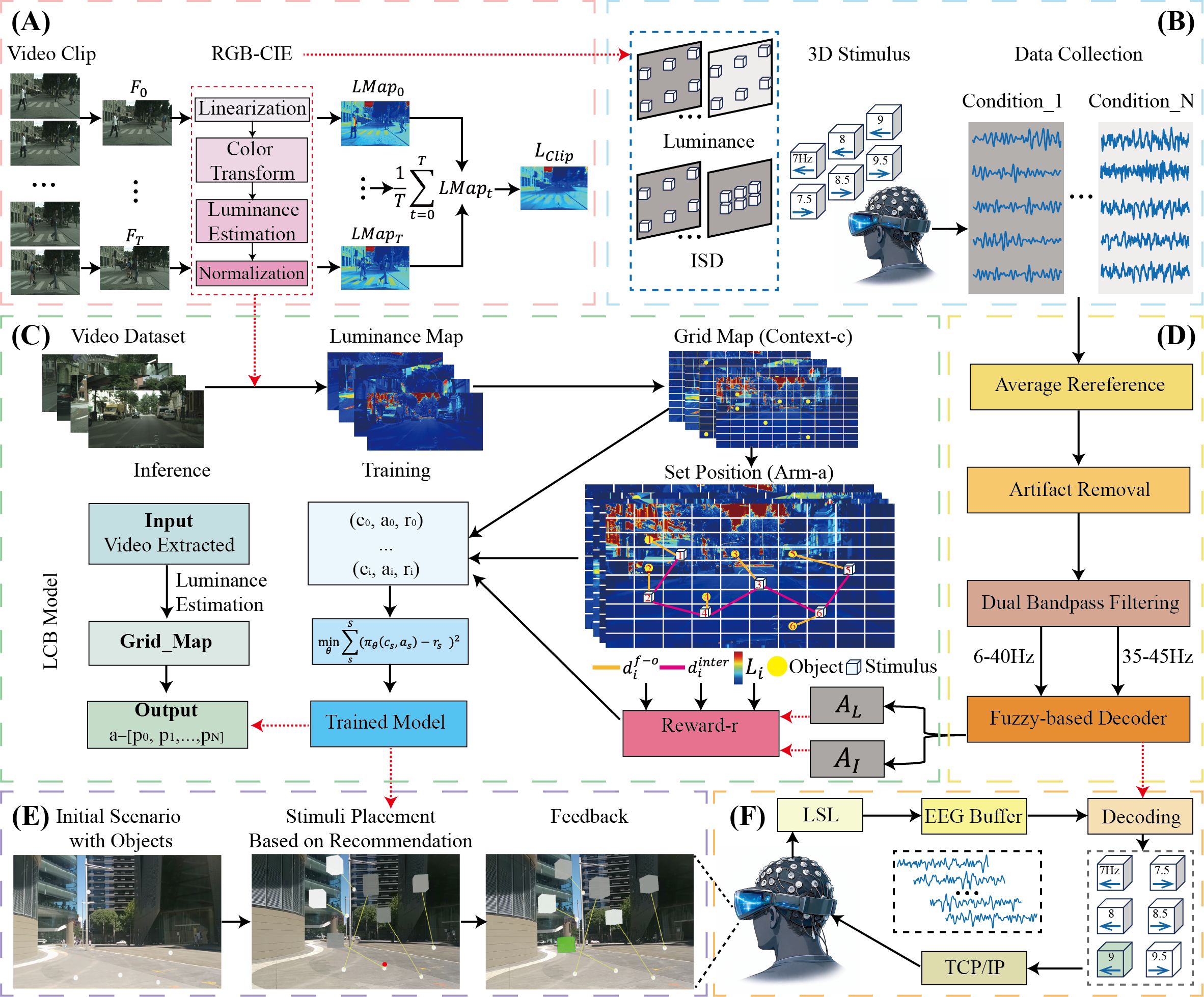}
    \caption{Overview of the SASLO system architecture and module interactions. In the diagram, solid black arrows represent data flow, whereas dashed red arrows denote invocation relationships between modules. \textbf{(A)} A scene luminance estimator. RGB frames ($F_0 \dots F_T$) are converted into luminance maps ($LMap_0 \dots LMap_T$) via RGB-CIE transformation and normalization, followed by temporal averaging across frames to produce the clip-level luminance map ($L_{\mathrm{Clip}}$). \textbf{(B)} Offline single-factor experiment for reward estimation, where SSVEP signals are elicited under systematically varied luminance and ISD conditions. Six 3D SSVEP stimuli are adopted, in which stimuli at 7, 8, and 9 Hz exhibit leftward rotation, while those at 7.5, 8.5, and 9.5 Hz exhibit rightward rotation. \textbf{(C)} LCB model for spatial layout optimization. During training, each sample is represented as a triplet $(c_t, a_t, r_t)$, where $c_t$ denotes the contextual luminance map, $a_t$ denotes the stimulus layout, and $r_t$ represents the corresponding reward. The training reward is constructed based on the offline decoding accuracies $A_L$ and $A_I$ obtained from the luminance-only and ISD-only experiments, respectively. The model learns to predict rewards for different spatial layouts based on contextual inputs. During inference, the extracted video is transformed into a grid context map with luminance estimation, and the trained LCB model outputs an optimized layout $a = \{p_0, p_1, \dots, p_n\}$ for stimulus placement. \textbf{(D)} EEG preprocessing and decoding pipeline. The preprocessing stage includes average re-referencing, artifact removal, and dual-band bandpass filtering. A dual-input fuzzy-based decoder then integrates frequency-related and rotation-related information extracted from two distinct frequency bands for SSVEP classification. \textbf{(E)} Illustration of the outdoor online AR single-trial testing scenario and experimental procedure. \textbf{(G)} Online AR-SSVEP system implementation, illustrating the real-time data flow.}
    \label{fig:Figure1}
\end{figure*}

\section{Methods}

\subsection{AR-SSVEP Paradigm}

\subsubsection{3D SSVEP Paradigm for AR-Based Object Selection}
In this work, a three-dimensional SSVEP (3D-SSVEP) paradigm proposed in our previous work is adopted \cite{cao2025novel}. In addition to frequency modulation, different rotation modes are integrated into the visual stimuli. Compared with conventional two-dimensional SSVEP (2D-SSVEP) stimuli that rely primarily on frequency modulation, the proposed 3D stimuli additionally elicit rotation-related EEG biomarkers. These complementary biomarkers provide extra discriminative features and thereby improve the robustness and decoding accuracy of AR-SSVEP.

As shown in Fig.~\ref{fig:Figure1}(B), six rotating 3D SSVEP stimuli are employed. The flickering frequencies of stimuli range from 7 Hz to 9.5 Hz with an increment of 0.5 Hz. Frequency modulation is implemented with a black-white sinusoidal waveform. Among the six stimuli, the stimuli flickering at 7 Hz, 8 Hz, and 9 Hz rotate in the leftward direction, while the remaining three stimuli rotate in the rightward direction. The rotation speed is set to 30 degrees per second. Each stimulus is designed with a fixed physical size of 20~cm $\times$ 20~cm. In the online experiment, the 3D-SSVEP paradigm is employed for an object selection task \cite{chen2020ssvep}, where each visual stimulus is associated with a corresponding virtual object in the AR environment, enabling participants to express their selection intention by fixating on the stimulus linked to the target object.
\subsubsection{Preprocessing Methods}

As shown in Fig.~\ref{fig:Figure1}(D), three essential preprocessing steps are applied in this work. First, an average reference is applied to the collected EEG to suppress baseline drift. Second, independent component analysis (ICA) is used to remove artifacts such as EOG and EMG \cite{zhukov2002independent}. Finally, following the method described in the previous study~\cite{cao2025novel}, FIR bandpass filters of 6-40 Hz and 35-45 Hz are applied to extract frequency- and rotation-related EEG components, respectively. After preprocessing, a dual-input fuzzy rule-based model is employed for 3D-SSVEP decoding.

\subsubsection{Decoding Model}

\begin{figure}[h!]
    \centering
    \includegraphics[width=1\linewidth]{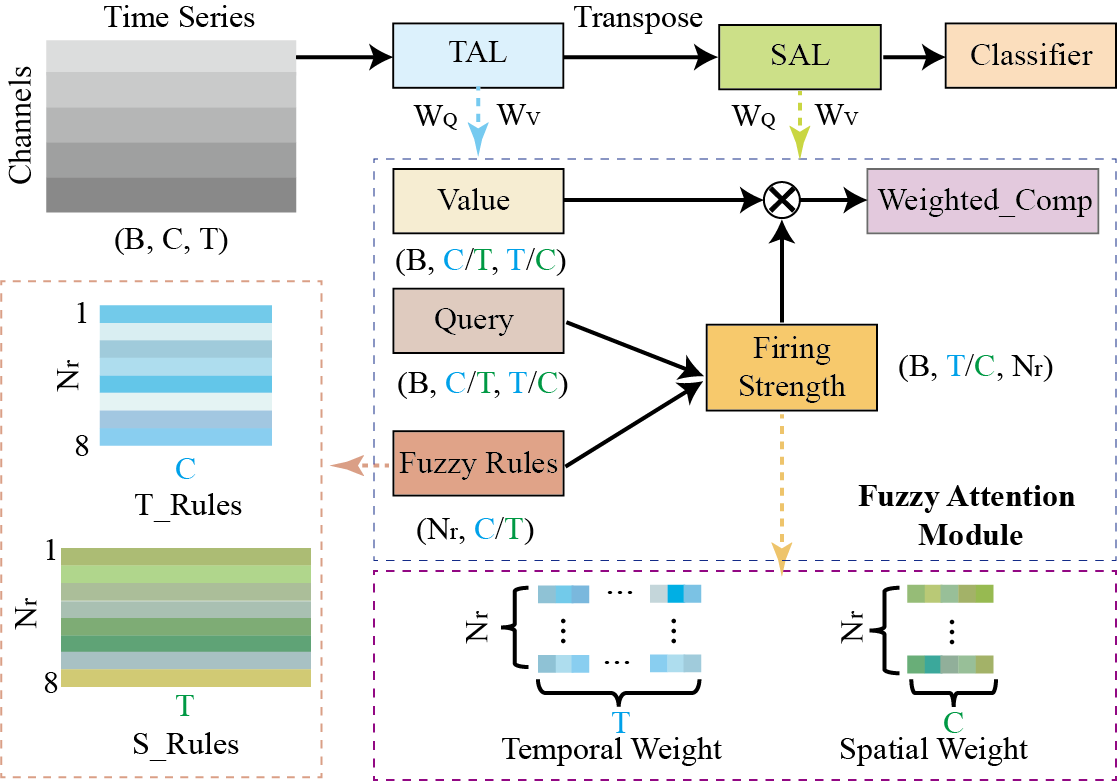}
    \caption{Diagram of the single-input fuzzy rule-based model. Blue and green elements denote the components of TAL and SAL, respectively.}
    \label{fig:Figure2}
\end{figure}
A fuzzy rule-based decoder is adopted to explicitly model data uncertainty while maintaining interpretability, which is critical for transparent and trustworthy BCI systems \cite{lin1996neural,jiang2025interpretable,cao2025emd}. In addition, fuzzy rule-based decoders have shown strong transferability in SSVEP decoding and are particularly effective under limited calibration conditions \cite{jiang2025ifuzzytl}. 
The architecture of the fuzzy decoder is illustrated in Fig.~\ref{fig:Figure2}, and further details can be found in \cite{jiang2025ifuzzytl}. The EEG input is represented as $(B, C, T)$, where $B$, $C$, and $T$ denote the batch size, number of channels, and signal length, respectively.

Following the fuzzy attention structure~\cite{jiang2025fuzzyatt}, the input EEG is first processed by a temporal attention layer (TAL), where the signals are linearly projected into query and value representations via learnable matrices $\mathbf{W}^{Q}$ and $\mathbf{W}^{V}$.
In TAL, $N_r$ temporal fuzzy rules, each associated with a learnable prototype vector, are initialized to model temporal dependencies. These rules are used to compute temporal firing strengths with shape $(B, T, N_r)$ via a Gaussian membership function (GMF) as Eq.~\ref{eq:eq1}, which are subsequently employed as temporal attention weights for the value representation.
\begin{equation}
\label{eq:eq1}
w_j(t)
=
\frac{
\exp\!\left(
-
\left\|
\mathbf{W}_j^{Q}\mathbf{x}(t) - \boldsymbol{r}_j
\right\|_{\boldsymbol{\Sigma}_j^{-1}}^{2}
\right)
}{
\sum\limits_{i=1}^{N_r}
\exp\!\left(
-
\left\|
\mathbf{W}_i^{Q}\mathbf{x}(t) - \boldsymbol{r}_i
\right\|_{\boldsymbol{\Sigma}_i^{-1}}^{2}
\right)
}
\end{equation}
with
\begin{equation}
\left\|
\mathbf{a}
\right\|_{\boldsymbol{\Sigma}^{-1}}^{2}
=
\mathbf{a}^{\top}
\boldsymbol{\Sigma}^{-1}
\mathbf{a}
\end{equation}
Here, $\mathbf{x}(t) \in \mathbb{R}^{C}$ denotes the EEG feature vector
across $C$ channels at time index $t$, and $w_j(t)$ represents the
normalized firing strength associated with the $j$-th fuzzy rule.
$\boldsymbol{r}_j \in \mathbb{R}^{C}$ and
$\boldsymbol{\Sigma}_j = \mathrm{diag}(\sigma_{j,1}^2, \dots,
\sigma_{j,C}^2)$ denote the learnable rule vector and its corresponding diagonal covariance matrix, respectively.
The temporal attention-weighted representation is obtained by aggregating value projections weighted by the attention coefficients.
\begin{equation}
\mathbf{y}(t)
=
\sum_{j=1}^{N_r}
w_j(t)\,
\mathbf{W}_j^{V}\mathbf{x}(t)
\end{equation}
where $\mathbf{y}(t) \in \mathbb{R}^{C}$ denotes the temporal attention-weighted EEG vector at time $t$.
Through this operation, informative temporal components in the EEG signal are enhanced, while irrelevant or noisy parts are suppressed.

The output of TAL is then transposed and fed into a
spatial attention layer (SAL).
The SAL shares the same structure as the TAL, except that attention is computed along the channel dimension, producing spatial attention weights with a shape of $(B, C, N_r)$. Finally, the weighted outputs of SAL are passed to a multilayer perceptron (MLP) classifier.
In this work, EEG signals from two frequency bands are processed in parallel by two fuzzy attention modules with identical structures. The outputs from the two SAL modules are then concatenated and fed into an MLP classifier.

\subsection{Luminance Estimation}

Luminance estimation is essential to the SASLO system. However, photometers and physical luminance measurements are not supported in most AR devices. Unlike offline illumination reconstruction tasks, the objective of this work is not to reconstruct physical lighting conditions, but to obtain a perceptual luminance descriptor associated with stimulus saliency and SSVEP performance. Therefore, an RGB-CIE-based luminance estimation method is adopted (Fig.~\ref{fig:Figure1}(A)), which calculates the perceptual luminance from RGB inputs using the CIE 1931 standard \cite{jung2024tailoring}. This method provides a robust and low-latency estimation of human-perceived luminance, ensuring compatibility with real-time AR systems.

As illustrated in Fig.~\ref{fig:Figure1}(A), the luminance extraction process comprises four sequential steps, including linearization, color transformation, luminance estimation, and normalization.
Given an input sRGB image denoted as
\begin{equation}
sRGB(x, y) = [R_s(x,y), G_s(x,y), B_s(x,y)],
\end{equation}
As shown in Eq.~\ref{eq:eq5}, the sRGB components $(R_s, G_s, B_s)$ are first converted into linear RGB values $(R, G, B)$ to compensate for gamma encoding and device-specific nonlinearities. 
\begin{equation}
C(x, y) = \left( \frac{C_s(x, y)}{255} \right)^{\gamma}, \quad C \in \{R, G, B\}
\label{eq:eq5}
\end{equation}
where $\gamma = 2.2$ corresponds to the effective gamma exponent of the standard sRGB color space \cite{epicoco2024can}.
 The linearized values are then transformed into the CIE 1931 XYZ color space using the standard colorimetric matrix \cite{jung2024tailoring}:
\begin{equation}
\begin{bmatrix} 
X(x,y) \\ 
Y(x,y) \\ 
Z(x,y)
\end{bmatrix}
=
\begin{bmatrix}
0.4124 & 0.3576 & 0.1805 \\
0.2126 & 0.7152 & 0.0722 \\
0.0193 & 0.1192 & 0.9505
\end{bmatrix}
\begin{bmatrix}
R(x,y) \\ 
G(x,y) \\ 
B(x,y)
\end{bmatrix}
\end{equation}
The luminance for each pixel is then computed from the CIE $Y$ channel \cite{jung2024tailoring}:
\begin{equation}
L(x,y) = 0.2126 R(x,y) + 0.7152 G(x,y) + 0.0722 B(x,y)
\end{equation}
which reflects the wavelength-dependent sensitivity of human vision and determines perceived luminance \cite{jung2024tailoring, fairchild2013color}. Finally, the luminance map is normalized to the range $[0,1]$ for consistency across devices and lighting conditions:
\begin{equation}
L_n(x,y) = \frac{L(x,y) - L_{\min}}{L_{\max} - L_{\min}}
\end{equation}
The resulting normalized map $L_n(x,y)$ forms a pixel-wise luminance heatmap, representing perceptual luminance distribution across the image. 
As illustrated in Fig.~\ref{fig:Figure1}(A), the recorded video consists of $T$ frames denoted as $\mathrm{F}_t$ ($t = 0, 1, \dots, T$), and a luminance map $\mathrm{LMap}_t$ is computed for each frame. The overall luminance distribution of the clip is then obtained by averaging all luminance maps as
\begin{equation}
    L_{\mathrm{Clip}} = \frac{1}{T} \sum_{t=0}^{T} \mathrm{LMap}_t
\end{equation}
In this work, a 1 s video clip was recorded at 30 fps, corresponding to $T = 30$ frames used for luminance estimation.

\subsection{LCB-Based Spatial Layout Recommendation}

In the SASLO system, another essential module is the recommendation model for optimizing the spatial layout of stimuli. LCB is commonly used in recommendation systems, especially for parameter tuning \cite{agrawal2016linear}. In this study, the problem of optimizing stimulus layout can be formulated as a recommendation task, where the luminance context extracted from the scene serves as the input, and the system recommends the optimal spatial layout of stimuli as the output. Therefore, the LCB model is adopted to optimize the stimulus layout, since it effectively handles recommendation problems driven by contextual features.

As shown in Fig.~\ref{fig:Figure1}(C), an LCB model consists of three elements: context ($c$), arm ($a$), and reward ($r$).
The objective of the LCB is to estimate the parameter $\boldsymbol{\theta}^*$ that governs the expected reward given the context $c_s$ and arm $a_s$, where $s$ denotes the index of the $s$-th sample. In the linear bandit setting, the observed reward $r_s$ is assumed to follow
\begin{equation}
r_s = \mathbf{x}_{s,a}^{\top}\boldsymbol{\theta}^* + \epsilon_s
\end{equation}
where $\mathbf{x}_{s,a}$ denotes the feature vector of the context--arm pair $(c_s,a_s)$, and $\epsilon_s$ is a zero-mean noise term. The parameter is estimated by minimizing the squared prediction error between the predicted and observed rewards:
\begin{equation}
\hat{\boldsymbol{\theta}} = \arg\min_{\boldsymbol{\theta}} 
\sum_s \left(\mathbf{x}_{s,a}^{\top}\boldsymbol{\theta} - r_s\right)^2
\end{equation}
During decision-making, the LCB adopts an upper confidence bound (UCB) strategy to balance exploration and exploitation:
\begin{equation}
a_s = \arg\max_a \left( \mathbf{x}_{s,a}^\top \hat{\boldsymbol{\theta}} + \lambda \|\mathbf{x}_{s,a}\|_{A^{-1}} \right)
\end{equation}
where $\lambda$ controls the exploration strength and $A^{-1}$ is the inverse covariance matrix accumulated from past observations.

In this work, as shown in the right part of Fig.~\ref{fig:Figure1}(C), the extracted luminance map is discretized into an $N_g \times N_g$ spatial grid, with each cell representing a potential position. Since SSVEP-based object selection is a classical and widely adopted application scenario, it is employed in this work as the experimental task. Accordingly, six simulated objects are randomly set on the luminance grid map to construct the context representation incorporating spatial object information. The final grid map serves as the input context for the LCB model. 

For a given context, an arm $a$ is defined as the complete configuration of stimulus positions:
\begin{equation}
a = (p_1, p_2, \dots, p_N)
\end{equation}
where $N$ denotes the number of stimuli and $p_i = p_i(x,y)$ denotes the position of the $i$-th stimulus. 
In this setting, the arm jointly encodes all stimulus positions, enabling the LCB model to optimize the overall spatial layout.

With regard to reward function, the SASLO system defines three components as follows, where $i$ denotes the $i$-th stimulus:
\begin{enumerate}
    \item the local luminance at each potential position \(L_i\);
\item the distance between each stimulus and its paired object, denoted as \(d_i^{(f\!-\!o)}\);
\item the ISD between a stimulus and its nearest neighbor, denoted as \(d_i^{(\mathrm{inter})}\).

\end{enumerate}
Among these reward components, both luminance- and ISD-related rewards are derived from offline single-factor experiments, from which empirical functions are fitted to model the nonlinear effects of these factors on performance.
For the evaluation of the luminance effect on performance, discrete background luminance levels $\{L_k\}_{k=1}^{K}$ are tested under a fixed ISD, and the measured SSVEP decoding accuracy $A_L$ is normalized to obtain an observed reward $r_L(L_k)$, which is further extended to a continuous function $\hat r_L(L)$ via interpolation.
Likewise, for the evaluation of the ISD effect, the background luminance is fixed while the ISD values $\{d_m^{\mathrm{inter}}\}_{m=1}^{M}$ are varied. 
The normalized ISD reward $r_D(d_m^{\mathrm{inter}})$ is derived from the experimental results $A_I$, and then interpolated to obtain a continuous function $\hat{r}_D(d^{\mathrm{inter}})$.
Detailed formulations of $\hat r_L(\cdot)$ and $\hat r_D(\cdot)$ are provided in Supplementary Sections~I.A and~I.B.

In addition, a stimulus-object distance (SOD) reward is introduced to promote an interaction-friendly spatial layout.
The parameter $d_i^{f\!-\!o}$ does not directly affect the SSVEP response but constrains each stimulus to remain visually aligned with its corresponding object.
Accordingly, SOD reward is defined as a normalized constraint:
\begin{equation}
\hat{r_O}_(d_i^{f\!-\!o}) =
\max\!\left(0,\; 1 - \frac{d_i^{f\!-\!o}}{d_{\max}}\right)
\end{equation}
where $d_{\max}$ denotes the maximum allowable distance between the stimulus and its corresponding object.

After normalization, the reward for a candidate layout $a$ is computed by aggregating luminance, ISD, and SOD rewards across all stimuli in a minimum-aware manner:
\begin{equation}
\begin{aligned}
r(a)
=
&\;\alpha \cdot \frac{1}{3N} \sum_{i=1}^{N}
\Big(
\hat r_L(L_i)
+
\hat r_D(d_i^{\mathrm{inter}})
+
\hat{r_O}(d_i^{f\!-\!o})
\Big) \\
&\;+
\frac{(1-\alpha)}{3}\cdot
\min_{i}
\Big(
\hat r_L(L_i)
+
\hat r_D(d_i^{\mathrm{inter}})
+
\hat{r_O}(d_i^{f\!-\!o})
\Big)
\end{aligned}
\end{equation}
where $\alpha \in [0,1]$ balances overall performance and worst-case robustness to prevent the optimization from being dominated by extreme cases.

During inference, the input scene is processed by the luminance estimator to obtain a discretized grid map, which serves as the contextual input.
Next, the trained LCB model evaluates a set of feasible stimulus layout candidates $\mathcal{A}$.
For each candidate arm, a layout-level reward $r(a)$ is predicted based on luminance, ISD, and SOD.
The optimal layout is then selected by maximizing the predicted reward.
\begin{equation}
a^{*} = \arg\max_{a \in \mathcal{A}} r(a)
\end{equation}

\subsection{Implementation of the LCB Model}
The LCB model was trained on the Cityscapes dataset \cite{cordts2016cityscapes}, which consisted of diverse street-scene video sequences collected from 50 cities and included pixel-level annotations for 5,000 RGB frames. All frames were resized from $2048 \times 1024$ to $1920 \times 1080$ to match the resolution of the video captured by the HoloLens~2 AR device (Microsoft, Redmond, WA, USA). The contextual input vector had 18 dimensions, formed by computing three features (luminance, ISD, and SOD) for each of the six stimulus-object pairs in the context-arm pair $(c,a)$. The regularization parameter $\alpha$ was set to 0.25, and the exploration coefficient $\lambda$ was set to 0.5, respectively. The parameter vector $\hat{{\theta}}$ was initialized to zero. The maximum SOD $d_{\max}$  was set to $3\sqrt{2}$ grid units, corresponding to a $3 \times 3$ local neighborhood around each object.

For inference, a stochastic sampling strategy was adopted. At each trial, 2000 candidate stimulus layouts were sampled from the arm space $\mathcal{A}$ under the same context, using a farthest-first strategy to enforce diversity and avoid excessive concentration in limited regions of the arm space \cite{basu2004active}. The trained LCB model was then used to predict the reward for each sampled arm, and the arm with the maximum score was selected once its predicted reward exceeded a threshold of 0.8. If no sampled arm satisfied this criterion, additional batches of candidates were generated until a valid arm was obtained. This threshold-based and diversity-aware sampling strategy enabled efficient real-time layout recommendation under computational constraints.

\section{Offline Experiments for Reward Construction}

\subsection{Experimental Objective}
Although previous studies have explored factors influencing AR-SSVEP \cite{zhang2022ambient, zhao2020ssvep, zhang2023improving}, these findings cannot be directly applied due to discrepancies in variable definitions and differences in hardware configurations. Consequently, directly transferring reward formulations from existing studies is not appropriate.

To approximate the underlying relationship between contextual factors and SSVEP performance in the SASLO system, the reward functions ($\hat{r}_L(\cdot)$ and $\hat{r}_D(\cdot)$) were derived from real human SSVEP data. Therefore, two single-factor offline simulation experiments were conducted to independently quantify the effects of luminance and ISD on AR-SSVEP performance.

\begin{figure}[h!]
    \centering
    \includegraphics[width=1\linewidth]{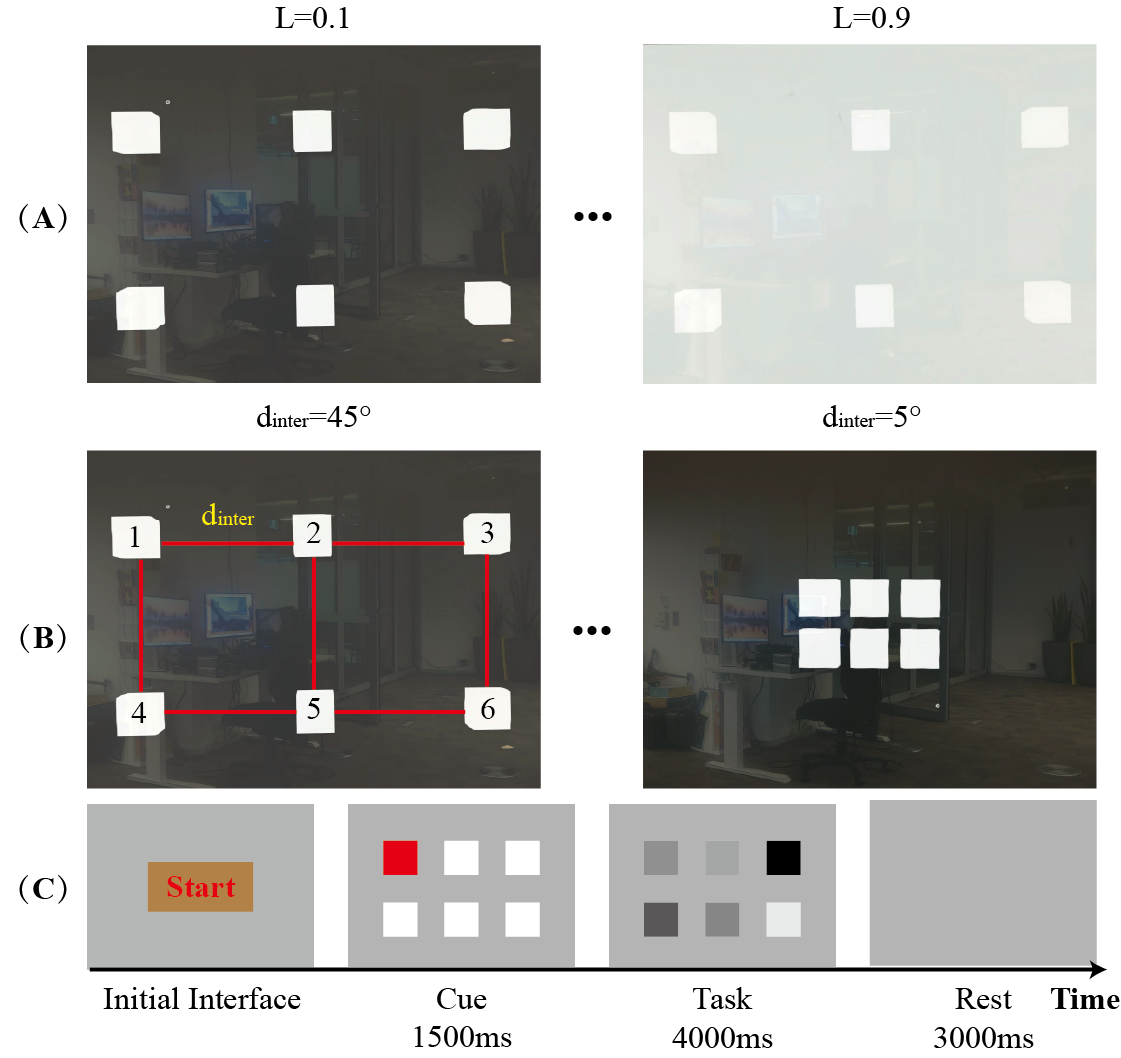}
    \caption{Illustration of the offline simulation paradigm and experimental procedure. \textbf{(A)} Luminance-based reward estimation experiment, where the stimulus luminance is systematically manipulated using normalized RGB-CIE luminance ($L$). \textbf{(B)} ISD-based reward estimation experiment, where the experimental variable is the ISD ($d_{inter}$). \textbf{(C)} Experiment procedure of a single trial.}
    \label{fig:Figure3}
\end{figure}

\subsection{Experimental Paradigm and Procedure}
The paradigms of the two single-factor experiments are illustrated in Fig.~\ref{fig:Figure3}(A) and (B). 
Since visual perception depends on retinal projection rather than physical distance \cite{foley2004visual}, the ISD $d_{\mathrm{inter}}$ was defined in terms of visual angle to ensure perceptual consistency across different viewing distances. In this study, the viewing distance was fixed at 3~m.
In the luminance experiment, the ISD of all flickers was fixed at $45^{\circ}$ of visual angle. Nine normalized CIE-Y luminance levels ranging from 0.1 to 0.9 with an increment of 0.1 were generated on the HoloLens~2. 
In the ISD experiment, the luminance was fixed at 0.1, and $d_{\mathrm{inter}}$ ranged from $5^{\circ}$ to $45^{\circ}$ in $5^{\circ}$ increments. The upper limit was set to $45^{\circ}$ as it approximately represents the widest horizontal field that can be perceived without head movement~\cite{jansen2010restricting}.

The procedure of a single trial is illustrated in Fig.~\ref{fig:Figure3}(C). At the beginning of the experiment, an initial interface with a virtual “Start” button was displayed. When subjects were ready, they could press the button to initiate a new round. Subsequently, a cue interface was presented for 1500~ms, during which the target stimulus was highlighted in red as a cue. Subjects were then instructed to fixate on the target stimulus for 4000~ms. Finally, a 3000~ms rest period was provided. Each round was completed when all stimuli had been selected once as the target. Each session was composed of nine rounds, with each round corresponding to one experimental condition. The order of conditions within each session was randomized. Each participant completed five sessions, with 270 trials conducted for each factor.

\subsection{Subjects and Experimental Setup}
Ten subjects (3 females and 7 males), aged between 21 and 33 years, participated in the offline experiment. All subjects were physically healthy and had normal or corrected-to-normal vision. Subjects were seated comfortably in a completely dark room to eliminate any external interference with the modulated luminance background. All subjects provided written informed consent (Grant number: UTS HREC REF No. ETH20-5371).

All experimental paradigms were presented with a HoloLens~2 headset. EEG signals were recorded with a 64-channel LiveAmp system (Brain Products GmbH, Gilching, Germany). The sampling rate was set to 500 Hz. Twelve electrodes (POz, PO8, PO7, PO6, PO5, PO4, PO3, PO1, PO2, O1, O2, and Oz) located over the occipital and posterior parietal regions were used for data collection, following the electrode configuration reported in \cite{cao2025novel}. At the onset of each task period, a digital marker corresponding to the target stimulus was transmitted to the recording server to facilitate data segmentation.

\subsection{Decoding Configuration}
As illustrated in Fig.~\ref{fig:Figure1}(D), the preprocessed signals were bandpass-filtered into two frequency bands (6-40 Hz and 35-45 Hz), corresponding to the frequency-related and rotation-related EEG components, respectively. The extracted two-band signals were then used as inputs to the fuzzy model~\cite{cao2025novel}. As illustrated in Fig.~\ref{fig:Figure2}, the number of adopted EEG channels $C$ was set to 12. The parameter $T$ was set to 1930, corresponding to 3.86 s at a sampling rate of 500 Hz. Although each trial lasted for 4 s, the initial 0.14 s were excluded to remove potential artifacts caused by eye movement \cite{wang2022stimulus}. The number of fuzzy rules $N_r$ was set to 5 in this work, which was determined empirically through multiple preliminary tests. The learning rate and dropout rate were set to 0.001 and 0.25 respectively. The Adam optimizer was employed for training \cite{kingma2014adam}, which was conducted on an NVIDIA GeForce RTX 4060 GPU (NVIDIA Corporation, Santa Clara, CA, USA).

\subsection{Results of Offline Experiments}

\begin{figure}[h!]
    \centering
    \includegraphics[width=1\linewidth]{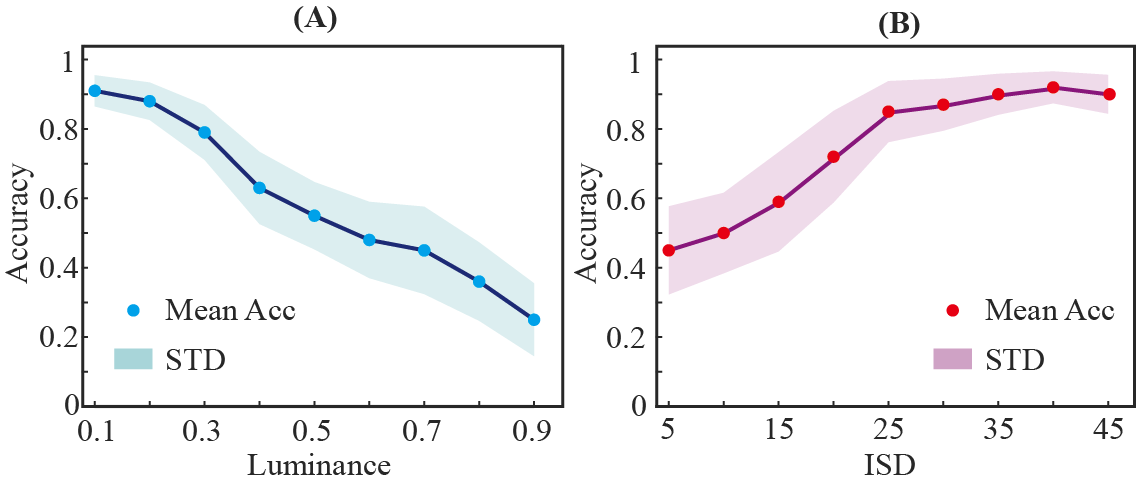}
    \caption{Results of the six-class offline experiment. \textbf{(A)} Variation of the average decoding accuracy with luminance levels. \textbf{(B)} Variation of the average decoding accuracy with ISD. Error bars represent the standard deviation (STD) across subjects}
    \label{fig:Figure4}
\end{figure}

The results of the two single-factor simulation experiments are illustrated in Fig.~\ref{fig:Figure4}. As shown in Fig.~\ref{fig:Figure4}(A), the decoding accuracy exhibits a negative correlation with luminance. The highest accuracy, denoted as $\mathrm{Acc}_{L_{\max}} = 91\%$, is achieved at a luminance level of 0.1, whereas the lowest accuracy, $\mathrm{Acc}_{L_{\min}} = 25\%$, is observed at a luminance level of 0.9. Specifically, as luminance increases from 0.1 to 0.4, the decoding accuracy exhibits an approximately quadratic decline. Within the range of 0.4-0.7, the accuracy decreases in an approximately linear manner. A steeper linear decline is observed as luminance further increases from 0.7 to 0.9. In contrast, decoding accuracy is positively correlated with the ISD, with the lowest accuracy of 46\% observed at $d_{\mathrm{inter}} = 5^{\circ}$ and the highest accuracy of 92\% achieved at $d_{\mathrm{inter}} = 40^{\circ}$. Notably, an approximately linear increase in accuracy is observed over the range of $5^{\circ}$-$25^{\circ}$, whereas no statistically significant improvement in accuracy is observed with further increases (pairwise comparisons between adjacent $d_{\mathrm{inter}}$, $p > 0.05$).

The offline experimental results indicate a monotonic dependence of decoding accuracy on both factors. However, the relationship between these factors and the true EEG reward cannot be well approximated by simple functions. Therefore, an interpolation-based reward is adopted in this work, as the relationship between the true EEG-derived reward and the factors can be more accurately approximated during LCB training. The detailed reward interpolation is provided in Supplementary Sections I.A and I.B.

\section{Outdoor Online Evaluation}
\begin{figure*}[h!]
    \centering
    \includegraphics[width=1\linewidth]{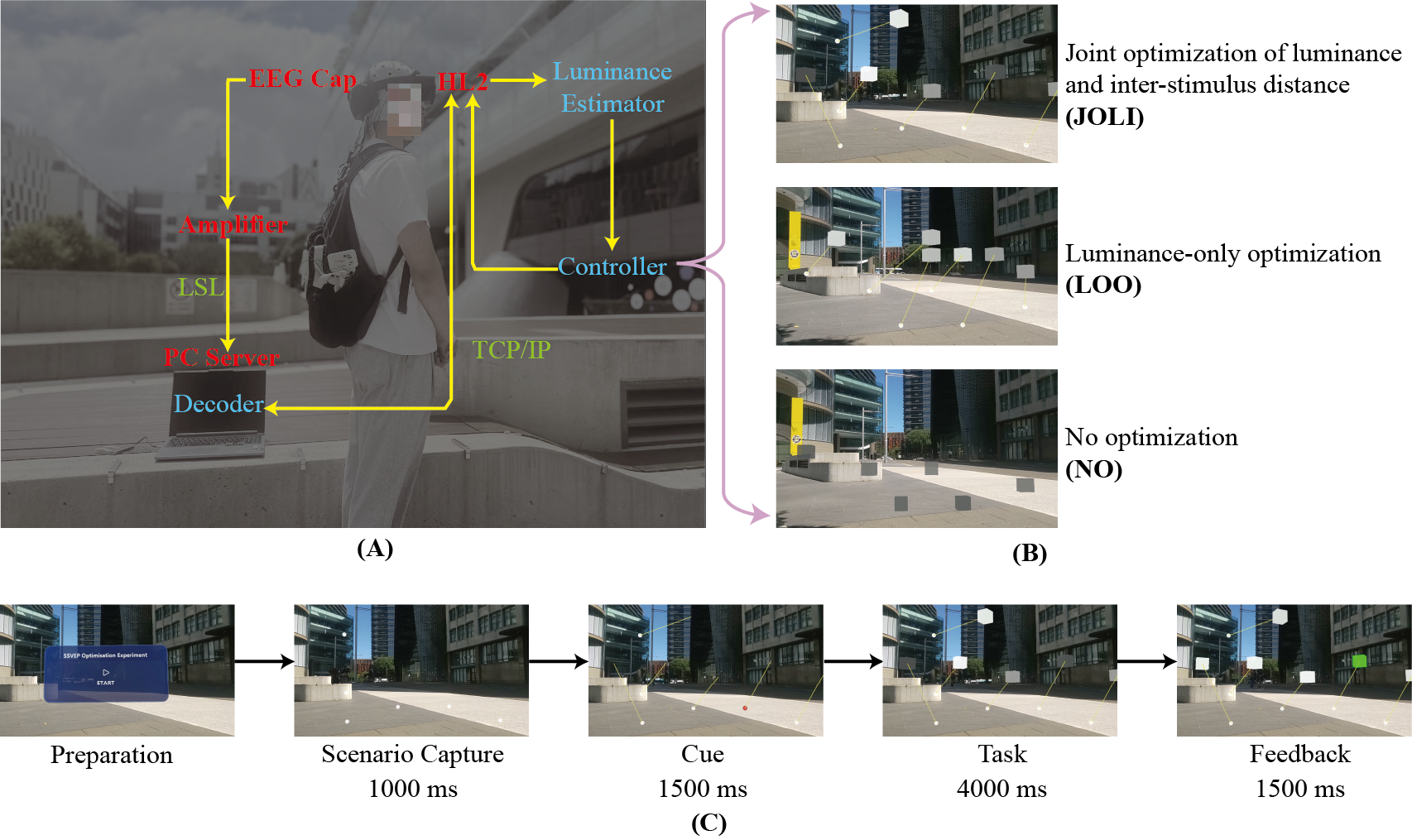}
    \caption{Overview of the outdoor online AR-SSVEP experiment. \textbf{(A)} Framework of the online system deployed in a real-world outdoor environment. Yellow arrows illustrate the data flow. Red bold labels denote hardware components. Blue labels indicate software modules. Green labels represent communication protocols. \textbf{(B)} Demonstration of stimulus spatial layouts under different optimization methods. \textbf{(C)} Single-trial online experimental procedure.}
    \label{fig:Figure5}
\end{figure*}

\subsection{Online System Architecture}

The architecture of the online system and the corresponding data flow are illustrated in Fig.~\ref{fig:Figure5}(A).
In the online experiment system, the luminance estimator and the LCB model were integrated into HoloLens~2 via Unity (Unity Technologies, San Francisco, CA, USA). The fuzzy decoder trained with offline data was deployed on a personal computer (PC) server. Two parallel threads were executed on the server. One thread handled real-time EEG acquisition, while a second thread listened for events from HoloLens~2 and performed SSVEP decoding.

For data acquisition, EEG signals were recorded from 12 channels, consistent with the offline setting. The acquired signals were transmitted to an 8-s circular buffer on the PC server through an amplifier via the Lab Streaming Layer (LSL) protocol. In each trial, a video clip was captured by the HoloLens~2, which was processed by the luminance estimator to generate a discretized luminance grid map with simulated object locations. The grid map was then fed into the LCB model to recommend an optimized stimulus layout. The stimuli in the recommended layout were subsequently presented to the subject. At the onset and offset of stimulus presentation, “start” and “end” digital events were transmitted from HoloLens~2 to the PC server via TCP/IP. These events were used to extract task-related EEG epochs from the buffer for SSVEP decoding. The decoding result was then transmitted back to HoloLens~2 to provide real-time feedback.

\subsection{Experimental Protocol}

\subsubsection{Task Design}
Following previous studies \cite{chen2020ssvep}, an SSVEP-based object selection task was adopted for outdoor online evaluation. In the online experiment, subjects were asked to select the target object by fixating on the corresponding stimulus, which was visually linked to the object by a yellow line (Fig.~\ref{fig:Figure5}(B)). To ensure the generalizability of the proposed SASLO system, six simulated virtual objects were randomly generated in the AR environment, as illustrated by the white spots in Fig.~\ref{fig:Figure5}(B) and (C). 

As shown in Fig.~\ref{fig:Figure5}(B), the proposed joint optimization of luminance and ISD (JOLI) method, was evaluated and compared with two baseline approaches. The first baseline was luminance-only optimization (LOO), in which the ISD factor was excluded from the reward and the LCB model was trained solely based on luminance. The second baseline followed previous work \cite{chen2020ssvep}, in which a non-optimization (NO) strategy was applied and the stimuli were directly overlaid onto their corresponding objects. 

\subsubsection{Experimental Procedure}
The detailed procedure of the online experiment is illustrated in Fig.~\ref{fig:Figure5}(C). At the beginning of the experiment, a preparation interface was presented, in which a virtual start button was provided for the participant. Following button activation, a round consisting of six trials was initiated.
In each trial, HoloLens~2 first captured the scene video for 1000~ms, and six virtual objects were randomly generated in the AR environment. A recommended stimulus layout was then generated by the LCB model based on the captured scene. Subsequently, one of the objects was highlighted in red as the target cue for 1500~ms. Next, all stimuli were presented for 4000~ms, during which the subject was instructed to fixate on the target stimulus. Finally, a feedback phase of 1500~ms was presented to indicate the predicted result (highlighted in green). Although the task duration was set to 4000 ms to mitigate potential instability at stimulus onset, real-time feedback was decoded using only the first 3000 ms of EEG data to evaluate user performance within a shorter time window. Owing to the strong transferability of fuzzy models, the decoder used for online experiment decoding was trained with data collected from offline experiments.

Within each round, the spatial layout of the objects was fixed after initialization, and the optimization method of layout remained unchanged. Each object was selected as the target once per round. A total of ten sessions were conducted, with each session comprising one round for each method (JOLI, LOO and NO). To mitigate potential fatigue effects, the order of the methods was randomized within each session.

\subsubsection{Subjects and Experimental Environment}
Subjects in the online experiment were the same as those in the offline study, and written ethical consent (Grant number: UTS HREC REF No. ETH20-5371) was obtained from all subjects. As illustrated in Fig.~\ref{fig:Figure5}(A), the outdoor experiment was conducted on a university campus between 8:00~am and 4:00~pm under sunny weather conditions. For ease of control, subjects were provided with an empty backpack to carry the portable amplifier and cables. During data collection, subjects were instructed to minimize body movements to reduce artifacts. 

\subsection{Online Experimental Results}

\begin{table*}[h!]
    \centering
    \captionsetup{width=\textwidth}
    \fontsize{8}{10}\selectfont
    \setlength{\tabcolsep}{3pt}

\caption{Average accuracy and ITR (bits/min) across subjects for three spatial layout optimization methods in the six-class outdoor online experiment under different input durations.
Asterisks indicate statistically significant differences compared with JOLI, determined by paired t-tests (*$p<0.05$, **$p<0.01$, ***$p<0.001$). Standard deviations are shown in parentheses. Bold values denote the best performance.}
\label{tab:table1}

\begin{tabular}{llllllll}
\toprule
\textbf{Metric} & \textbf{Method} & \textbf{1.0 s} & \textbf{1.5 s} & \textbf{2.0 s} & \textbf{2.5 s} & \textbf{3.0 s} & \textbf{3.5 s} \\
\midrule

\multirow{3}{*}{ACC}
& JOLI & \textbf{0.41 (0.11)} & \textbf{0.59 (0.11)} & \textbf{0.70 (0.12)} & \textbf{0.82 (0.08)} & \textbf{0.89 (0.09)} & \textbf{0.91 (0.06)}\\
& LOO & 0.30 (0.08)** & 0.51 (0.10)** & 0.67 (0.10)  & 0.77 (0.13)* & 0.85 (0.09)* & 0.88 (0.08)* \\
& NO & 0.29 (0.08)*** & 0.39 (0.06)*** & 0.52 (0.09)*** & 0.64 (0.08)*** & 0.75 (0.06)*** & 0.80 (0.07)*** \\

\midrule

\multirow{3}{*}{ITR}
& JOLI & \textbf{14.12 (11.43)} & \textbf{25.03 (11.83)} & \textbf{30.20 (13.11)} & \textbf{35.02 (9.05)} & \textbf{35.74 (8.42)} & \textbf{32.78 (5.51)} \\
& LOO & 5.15 (5.11)** & 17.66 (9.27)** & 26.25 (9.26)* & 30.10 (11.24)* & 31.68 (8.19)* & 30.22 (6.73) \\
& NO & 4.62 (3.35)** & 7.60 (3.83)** & 13.59 (5.84)*** & 18.87 (6.18)*** & 22.89 (4.85)*** & 23.69 (4.91)*** \\

\bottomrule
\end{tabular}
\end{table*}

To evaluate online performance, classification accuracy and information transfer rate (ITR) are reported.
ITR serves as a comprehensive metric for online BCI performance by jointly accounting for classification accuracy, the number of stimuli, and the decoding time \cite{jiang2025ifuzzytl}. 
Therefore, ITR provides a more practical measure of system efficiency and usability in real-time interaction scenarios.
The ITR is calculated as
\begin{equation}
\mathrm{ITR} = \left[ \log_2 N_t + P \log_2 P + (1 - P) \log_2 \left( \frac{1 - P}{N_t - 1} \right) \right] \times \frac{60}{T_c},
\end{equation}
where $N_t$ denotes the number of stimuli ($N_t = 6$ in this work), $P$ represents the classification accuracy, and $T_c$ denotes the time cost for single-trial decoding. 
Specifically, $T_c$ is defined as the sum of the decoder input duration and an additional attention-shift duration of 0.135~s.
For metric calculation, the input data were strictly obtained from the online buffer and corresponded to the task phase of each trial.

The results of the online experiment are summarized in Table~\ref{tab:table1}. 
Classification accuracy consistently improves with increasing input duration for all methods.
Notably, both the accuracy and ITR achieved by JOLI are significantly higher than those of LOO and NO in most cases. 
With a 3~s input duration, JOLI achieves an accuracy of 0.89, whereas the corresponding accuracies are 0.85 for LOO and 0.75 for NO.
In terms of ITR, the highest values for both JOLI and LOO are obtained at 3~s, reaching 35.74~bits/min and 31.68~bits/min, respectively. By comparison, NO achieves its maximum ITR of 23.69~bits/min at a longer input duration of 3.5~s. 
These results indicate that scene luminance has a substantial impact on outdoor AR-SSVEP performance.
When isolating the effect of ISD, JOLI exhibits a more significant advantage over LOO under shorter time windows. 
For instance, with input durations of 1~s and 1.5~s, JOLI achieves accuracies of 0.41 and 0.59, respectively, whereas LOO reaches only 0.30 and 0.51. 
As the time window increases, the performance gap between JOLI and LOO gradually narrows, with LOO demonstrating relatively ideal performance.
This observation suggests that spatially adjacent stimuli induce stronger attentional interference in the early stage of SSVEP, whereas prolonged stimulation promotes attentional stabilization.

Individual accuracy is reported in Supplementary Table~I. 
When the input duration exceeds 3~s, eight subjects achieve classification accuracies higher than 0.8 with JOLI method. 
Notably, Subject~1 achieves an accuracy of 100\% with a 3.5~s input window. 
Across all subjects, JOLI consistently outperforms the NO method.
For shorter input durations ($\leq$1.5~s), JOLI yields higher accuracy than LOO for all subjects except Subject~6 and 9. 
However, when the input duration exceeds 2~s, several subjects, such as 2, 5, 7 and 8, achieve comparable or even higher accuracies with the LOO method.

\section{Discussion}
\subsection{Analysis of Biomarkers across Methods}

\begin{figure*}[h!]
    \centering
    \includegraphics[width=1\linewidth]{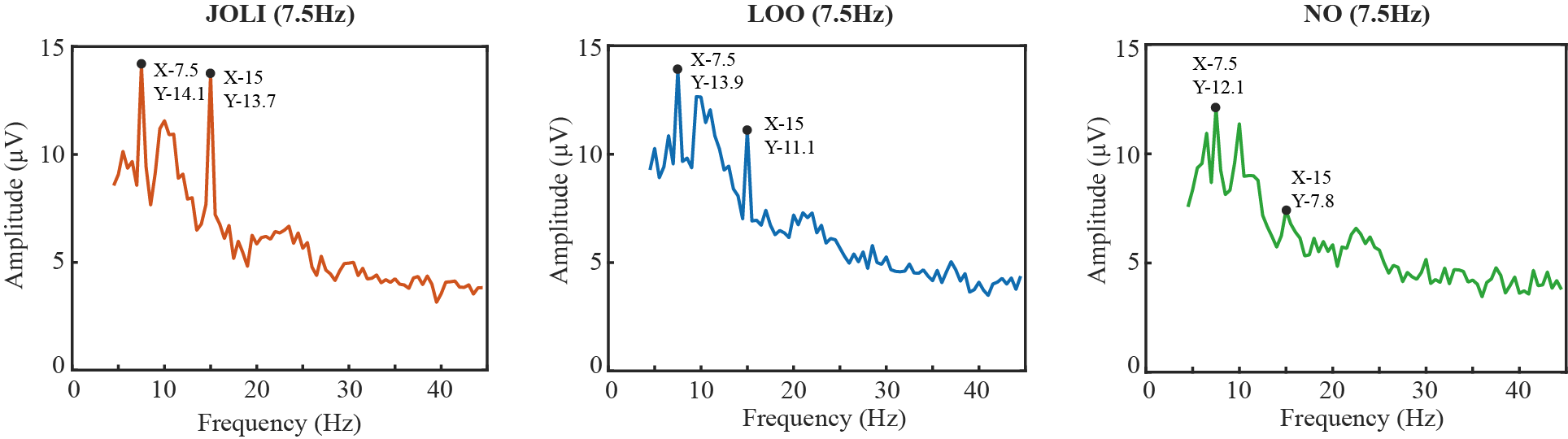}
    \caption{FFT-based amplitude spectra at 7.5 Hz across different methods. Single-sided FFT spectra averaged across subjects are shown for JOLI, LOO, and NO. Peaks at the target frequency (7.5 Hz) and its first harmonic (15 Hz) are labeled.}
    \label{fig:Figure7}
\end{figure*}

To investigate the effects of different methods on AR-SSVEP performance, the strength of task-related biomarkers was visualized.
In this work, a 3D-SSVEP paradigm was employed to elicit both the frequency-related biomarker (FRB) and the rotation-related biomarker (RRB).

Specifically, the FRB reflects neural responses synchronized with the fundamental and harmonic frequency of the stimulus. Therefore, FFT-based spectral analysis was performed on EEG signals from the Oz channel, which is strongly associated with frequency synchronization in SSVEP paradigms \cite{wong2021transferring}. The complete amplitude spectra for all stimulus classes are presented in Supplementary Fig.~1, and a representative example for the 7.5 Hz case is illustrated in Fig.~\ref{fig:Figure7}. As shown in Fig.~\ref{fig:Figure7}, the amplitudes of both the fundamental frequency and the first harmonic under the JOLI method are higher than those under LOO and NO. While the difference at the fundamental frequency is not statistically significant ($p > 0.05$), a significant difference is observed at the first harmonic when comparing NO with the other two methods ($p_{\text{JOLI-NO}} = 0.009$, $p_{\text{LOO-NO}} = 0.028$, $\alpha=0.05$). Consistent trends are also observed across other classes, as shown in Supplementary Fig.~1. Compared with JOLI and LOO, the task-related frequency peaks (7.5 Hz and 15 Hz) under the NO condition are less distinguishable from background components, resulting in a lower signal-to-noise ratio (SNR). These findings suggest that luminance interference may suppress the elicitation of the first harmonic response, which is consistent with previous studies~\cite{zhang2022ambient,zhang2023improving}. In contrast, the clearer fundamental and harmonic peaks observed with the JOLI method indicate that optimizing the spatial layout of stimuli by jointly considering luminance and ISD can enhance frequency synchronization, thereby improving AR-SSVEP performance.

\begin{figure}[h!]
    \centering
    \includegraphics[width=0.8\linewidth]{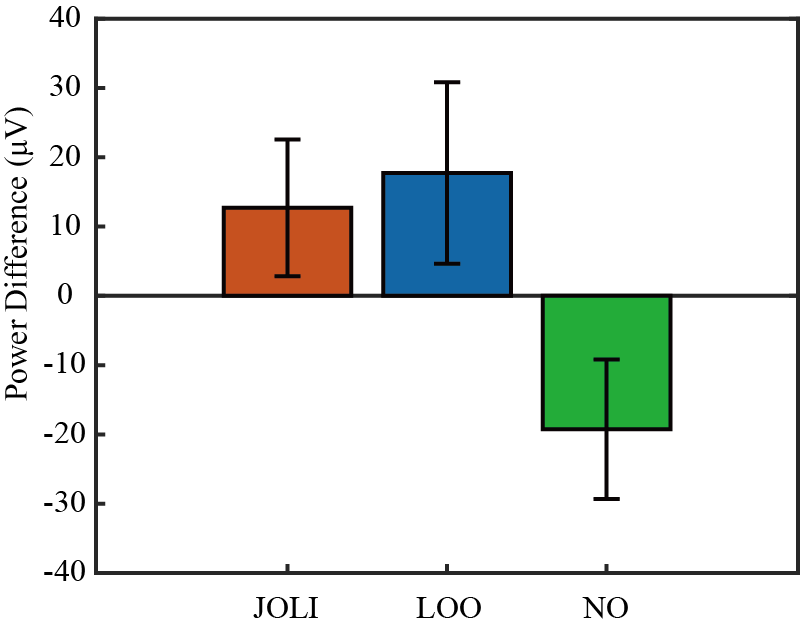}
    \caption{Comparison of rotation-related power differences under different methods. The mean EEG power differences between left- and right-rotation conditions at the PO5 channel are shown for JOLI, LOO, and NO. Error bars denote standard deviations across subjects.}
    \label{fig:Figure8}
\end{figure}

With respect to the RRB analysis, previous studies on 3D-SSVEP have demonstrated that visual rotation enhances low gamma-band (30-45 Hz) EEG activity in the parieto-occipital region ipsilateral to the rotation direction, with a more pronounced effect in the left hemisphere \cite{cao2025novel}. Therefore, low gamma-band EEG power was computed from the PO5 channel, which is located at the junction of the left occipital and parietal cortices. The power difference between left- and right-rotation EEG signals was adopted as the observation metric for RRB quantification. Under ideal elicitation conditions, a clear positive power difference between left- and right-rotation conditions is expected \cite{cao2025novel}.
The results of the rotation-related power difference are shown in Fig.~\ref{fig:Figure8}. For both JOLI and LOO, a clear positive power difference is observed at the PO5 channel, indicating effective elicitation of the RRB. Notably, the power difference under the LOO method exceeds that under JOLI, suggesting that rotation-related neural responses are relatively insensitive to ISD. In contrast, a negative power difference is observed under the NO method, indicating that increased luminance interference suppresses the reliable elicitation of the RRB, thereby leading to degraded SSVEP decoding performance.

\subsection{Interpretability Analysis of Fuzzy Decoding}

According to previous studies, the firing strength of the TAL can be employed to interpret the behavior of the fuzzy model, as it has been shown to preserve the temporal structure of the input signal \cite{jiang2025ifuzzytl, cao2025emd}. 
The firing strength reflects how the fuzzy model attends to and interprets the EEG signal during inference. After training, the temporally re-projected firing strength is expected to encode frequency-related information consistent with the input EEG. To support interpretability and explain the observed performance differences among the three methods, we analyzed the temporal firing strength of the fuzzy model.

\begin{figure}[h!]
    \centering
    \includegraphics[width=1\linewidth]{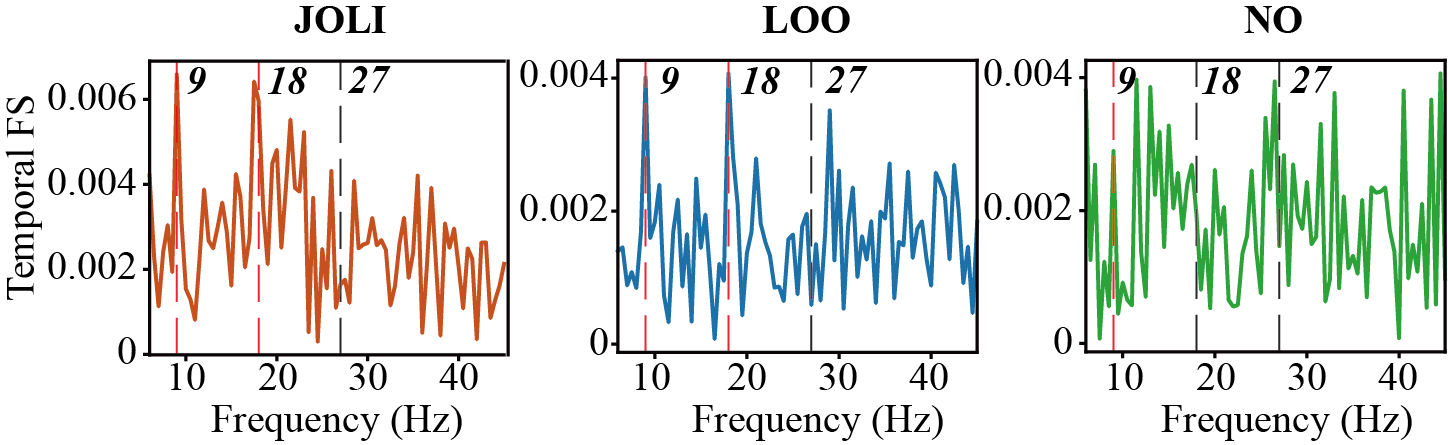 }
    \caption{Spectra of temporal firing strength inferred by the fuzzy model under the JOLI, LOO, and NO methods for the 9 Hz stimulus in the range of 6-45 Hz. Reference baselines for the fundamental frequency, the first harmonic, and the second harmonic are indicated by dashed lines. Red dashed lines denote baselines aligned with the task-related spectral peaks, whereas black dashed lines indicate misalignment.}
    \label{fig:figure9}
\end{figure}

Specifically, the frequency content of the temporal firing strength was examined using FFT, averaged within each class for each subject, and then averaged across ten subjects. This analysis provides a human-understandable representation of the learned knowledge, which is particularly relevant for SSVEP tasks. The resulting spectra are shown in Supplementary Fig. 2, with a representative 9 Hz example presented in Fig. \ref{fig:figure9}.

The firing strength reflects how the fuzzy model attends to and interprets the EEG signal during inference. After training, the temporally re-projected firing strength is expected to encode frequency-related information consistent with the input EEG. To support interpretability and explain the observed performance differences among the three methods, we analyzed the temporal firing strength of the fuzzy model.

Specifically, the frequency content of the temporal firing strength was examined using FFT, averaged within each condition for each subject, and then averaged across ten subjects. This analysis provides a human-understandable representation of the learned knowledge, which is particularly relevant for SSVEP tasks. The resulting spectra are shown in Supplementary Fig. 2, with a representative 9 Hz example presented in Fig. \ref{fig:figure9}.

In Fig.~\ref{fig:figure9}, clear alignment between the temporal firing strength spectra and the fundamental as well as first harmonic frequencies is observed under JOLI and LOO. Prominent peaks at 9 Hz and 18 Hz indicate that discriminative frequency information is effectively captured by the fuzzy model.
Under the NO method, only a weak fundamental frequency peak is present and is largely masked by noise-related spectral components, suggesting insufficient learning of frequency-specific features.

Similar patterns are observed in Supplementary Fig.~2. JOLI consistently produces distinct peaks at the fundamental and harmonic frequencies across all classes, while LOO shows pronounced task-related peaks for most classes except 7.5 Hz. In contrast, NO exhibits weak or suppressed task-related peaks.
These observations indicate that stimulus optimization methods substantially influence the signal quality of SSVEP, which in turn affects frequency learning in the fuzzy model and leads to performance differences.

\section{Conclusions and Future Work}

In this work, a SASLO system for AR-SSVEP is proposed to adaptively optimize multi-stimulus layouts based on JOLI. Specifically, an RGB-CIE-based luminance estimation method is used to estimate perceptual scene luminance, and the extracted context is fed into an LCB model for automatic optimization of AR-SSVEP stimulus layouts. Two pilot single-factor offline experiments were conducted to investigate the effects of luminance and ISD on AR-SSVEP performance, providing a basis for constructing reliable reward functions for LCB training. Furthermore, an outdoor online experiment involving ten subjects was performed to compare the proposed JOLI method with two baselines, including LOO and NO. With JOLI, an online classification accuracy of 0.89 and an ITR of 35.74~bits/min are achieved using a 3 s input window.
These results demonstrate a significant performance improvement achieved by the proposed SASLO system, especially under short time windows. 

While the proposed JOLI-based SASLO system achieves substantial performance improvements in outdoor AR-SSVEP scenarios, several challenges remain to be addressed. First, the current system optimizes stimulus layouts on a two-dimensional plane, whereas real-world AR environments are inherently three-dimensional. Depth information should be incorporated to enable more realistic and effective spatial layout optimization for AR-SSVEP stimuli. Second, this work focuses on only two influencing factors. Factors such as stimulus size and background motion intensity are also known to affect SSVEP responses and should be considered for online adaptive optimization. Future studies will therefore explore a broader range of scene- and stimulus-related factors to enhance the robustness of AR-SSVEP systems.

\bibliographystyle{IEEEtran}
\bibliography{reference}
\end{document}